\documentclass[11pt,a4paper]{article}
\usepackage[T1]{fontenc}
\usepackage{lmodern,amsmath,amssymb,amsthm,mathtools}
\usepackage{microtype}
\usepackage[margin=25mm]{geometry}
\usepackage{enumitem,booktabs,array}
\usepackage[colorlinks=true,allcolors=blue]{hyperref}
\hypersetup{pdftitle={A 17/7 Circuit Lower Bound for Binary-Encoded 3-SAT},pdfauthor={Theodoros Yianni and Sahibzada Allahyar},pdfsubject={Boolean circuit complexity},pdfkeywords={3-SAT, circuit lower bounds, gate elimination, formal verification}}
\newtheorem{theorem}{Theorem}[section]
\newtheorem{lemma}[theorem]{Lemma}
\newtheorem{proposition}[theorem]{Proposition}
\theoremstyle{definition}
\newtheorem{definition}[theorem]{Definition}
\newcommand{\bits}{\{0,1\}}
\newcommand{\C}{C_{B_2}}
\newcommand{\degF}{\deg_{\mathbb F_2}}
\newcommand{\succS}{\operatorname{succ}}
\newenvironment{proofsteps}{\begin{enumerate}[label=\textup{\arabic*.},ref=\arabic*,leftmargin=*,itemsep=4pt,topsep=4pt]}{\end{enumerate}}
\title{A $(17/7-o(1))n$ Circuit Lower Bound\\for Binary-Encoded 3-SAT}
\author{Theodoros Yianni\\
\small Department of Computer Science\\[-2pt]\small Royal Holloway, University of London
\and Sahibzada Allahyar\\\small\texttt{smaa4@cantab.ac.uk}}
\date{}
\begin{document}
\maketitle
\begin{abstract}
We strengthen the elementary $n-1$ circuit lower bound for a specified binary encoding of 3-SAT to
$(17/7)n-O(n\log\log n/\log n)$ gates over the full two-input Boolean basis, with unrestricted depth and sharing.
For each integer $w\ge5$, we encode 3-CNF formulas with exactly $m=2^w-1$ clauses over $m$ available variables using $n=3m(w+1)$ bits.
The proof embeds Boolean tests in variable identifiers while retaining $n-o(n)$ free bits, and combines weighted gate elimination with a nonlinear-gate lower bound from the tests' full algebraic degree over $\mathbb F_2$.
We provide Lean proofs of the finite inequality and the asymptotic bound for arbitrary acyclic binary circuits.
\end{abstract}
\section{Introduction}\label{sec:introduction}
We measure the circuit complexity of 3-SAT against the length of its complete binary formula description.
The NP-completeness of 3-SAT~\cite{Trevisan3} motivates seeking such bounds, but preserving a leading constant requires control of the encoding length.
If a $t$-bit input is reduced to a $t^2$-bit formula description, a bound linear in $t$ is only of square-root order in the description length.
We embed hard Boolean tests directly in variable identifiers, retaining $n-o(n)$ of the $n$ description bits to obtain a lower bound with leading coefficient $17/7$ over the full binary basis.

\paragraph{Main result.}
For an integer $w\ge5$, let $m=2^w-1$ and let $F_w$ be the 3-SAT decision function for formulas with exactly $m$ three-literal clauses over $m$ available variables.
We encode each literal using a $w$-bit variable identifier and a sign bit, giving $n=3m(w+1)$ input bits.
In the encoding of Section~\ref{sec:setting}, identifier zero is rejected, while unused variables, repeated literals, repeated clauses and tautologies are permitted.
Write $\C(f)$ for the minimum number of internal gates in an acyclic circuit computing $f$, with arbitrary Boolean functions of at most two inputs and unrestricted depth and sharing.

\begin{theorem}[Main result]\label{thm:overview}
Along the input lengths $n=3(2^w-1)(w+1)$ as $w\to\infty$,
\[
\C(F_w)\ge \frac{17}{7}n-O\!\left(\frac{n\log\log n}{\log n}\right),
\]
and a restriction based on a weighted sum modulo a prime gives
$\C(F_w)\ge(17/7)n-O(n/\sqrt{\log n})$.
\end{theorem}
Theorem~\ref{thm:main} gives the finite inequality and proves both estimates.
The bound is measured against all description bits, including identifiers and signs, and strengthens the elementary $n-1$ bound for this encoding.

\paragraph{Related work.}
Gate elimination dates to Schnorr~\cite{Sch74} and underlies a sequence of linear lower bounds for particular Boolean functions, including coefficients of $5/2$ due to Stockmeyer and Paul~\cite{Sto77,Pau77} and $3$ due to Blum~\cite{Blu84}.
More than thirty years after Blum's result, Find, Golovnev, Hirsch and Kulikov exceeded the coefficient $3$~\cite{FGHK16}; Li and Yang subsequently obtained $3.1n-o(n)$~\cite{LY22}.
Mishra and Santhanam give $4n-O(\log n)$ for a function in $\mathsf P$~\cite{MS26}, but transferring such a bound to $F_w$ requires control of both the reduction size and its output length.

Lachish and Raz introduced strong two-dependence, which keeps the four functions obtained by fixing a pair of inputs distinct after many other inputs have been fixed; they used it to prove a lower bound over the basis excluding XOR and its negation~\cite{LR01}.
Zwick's work examines fan-out over a restricted basis~\cite{Zwi91}, while weighted gate elimination accounts separately for the costs of different gate types~\cite{GKST16}.
Kojevnikov and Kulikov combine weighted elimination with an algebraic-degree bound on nonlinear gates~\cite[Section~3]{KK10}; our gate count uses this combination with two levels of input credit.

\paragraph{Proof strategy.}
The embedding uses clauses with three copies of one negative literal to force selected variables false.
Because the remaining clauses contain only positive literals, setting every unforced variable true satisfies them exactly when each clause contains a variable not forced false.
The free identifier bits therefore feed separate copies of a Boolean test for whether the identifier names an unforced variable.
A suitable choice of the number of free bits per identifier retains $n-o(n)$ bits; a transfer lemma also counts the gates removed with the fixed bits (Sections~\ref{sec:embedding}--\ref{sec:transfer}).

We choose Boolean tests with two properties that support the gate count.
Whenever sufficiently many bits remain free, the four assignments to any pair of them give four distinct functions of the other free bits, allowing gate elimination to continue as more inputs are fixed.
The tests also have full algebraic degree, the largest possible degree for their number of inputs, which forces additional nonlinear gates.
We obtain these tests by a counting argument and a weighted modular construction (Section~\ref{sec:blocks}).

The gate count assigns weight seven to parity gates (XOR and its negation) and weight four to nonlinear gates, then subtracts up to two credits per input.
An input receives one credit when its sole successor is nonlinear, and a second when that gate's sole successor is also nonlinear.
At each elimination step, an input can be fixed to decrease this quantity by at least fourteen; in the tight case, removed gates supply twelve units and two new credits supply the other two.
Before elimination, the embedded function has full degree on its $L$ free inputs, so Schnorr's degree bound forces at least $L-1$ nonlinear gates~\cite{Sch89}.
Raising their weight from four to seven supplies a further $3L-3$ units while making every gate's weight seven.
Summing the fourteen-unit decreases and adding this degree bound yields the coefficient $17/7$, with lower-order losses from the inputs left in each block when elimination stops.
Sections~\ref{sec:accounting}--\ref{sec:elimination} establish the local accounting, and Section~\ref{sec:main} sums it and transfers the bound to the complete SAT description.

\paragraph{Formal verification.}
The Lean formalization, archived on Zenodo~\cite{Companion}, verifies the finite bound and both asymptotic estimates for arbitrary acyclic binary circuits.
It relates acyclic circuit graphs with shared predecessors to the ordered representation used for elimination.
Section~\ref{sec:formalization} describes the paper-to-code correspondence, parameter hypotheses and reproducibility checks.

\section{Encoding and circuit model}\label{sec:setting}
Throughout, $\bits=\{0,1\}$ and all logarithms have base two.

\begin{definition}[Binary encoding of 3-SAT]\label{def:param}\label{def:lit}\label{def:enc}\label{def:F}
Fix an integer $w\ge5$ and set $m=2^w-1$.
A formula has exactly $m$ ordered clauses over the available variables $X_1,\ldots,X_m$.
Each clause is encoded by three consecutive literal fields in literal order.
A field consists of identifier bits $b_0,\ldots,b_{w-1}$ followed by a sign bit $s$, with identifier
$v=\sum_{j=0}^{w-1}2^j b_j$.
For $1\le v\le m$, the field denotes $X_v$ when $s=0$ and $\neg X_v$ when $s=1$; identifier zero is invalid.
Concatenating the clause records gives $n=3m(w+1)$ bits.
Variables may be absent, and repeated literals, repeated clauses and tautologies are allowed.

The total Boolean function $F_w:\bits^n\to\bits$ is zero if any identifier is invalid, and otherwise is one exactly when the encoded formula is satisfiable.
\end{definition}

\begin{definition}[Circuit model and complexity]\label{def:C}\label{def:size}\label{def:CC}
A $B_2$ circuit is a finite acyclic network with labelled primary inputs and a designated output.
Each gate computes an arbitrary Boolean function of at most two predecessor values; a predecessor may supply both arguments.
Depth and sharing are unrestricted, and circuits at different input lengths may be chosen independently.
The size $|C|$ is the number of internal gates, including unary and constant gates; inputs and wires contribute zero.
The circuit complexity $\C(f)$ is the minimum size of a $B_2$ circuit computing $f$ on every input.
\end{definition}
Acyclicity lets us order gates after their predecessors; we use this order to simplify circuits under restrictions.

\begin{definition}[Dependence and restrictions]\label{def:ess}\label{def:res}\label{def:cof}
An input is \emph{essential} if two assignments differing only at that input can have different outputs.
A constant restriction $\rho$ fixes specified input bits to specified values.
The resulting function on the remaining inputs is the \emph{cofactor} $f|_\rho$.
Fixing a pair of bits gives four cofactors, some of which may coincide.
\end{definition}

\begin{lemma}[Elementary size bound]\label{lem:floor}
A $B_2$ circuit with $N\ge1$ essential inputs has at least $N-1$ gates.
\end{lemma}
\begin{proof}
Delete all vertices with no path to the output, retaining every essential input.
If the remaining circuit has $a\ge N$ input vertices and $s$ gates, its underlying undirected graph is connected and has at least $a+s-1$ edges.
At most $2s$ wires enter the gates, so $a+s-1\le2s$ and hence $s\ge N-1$.
Deletion did not increase the gate count.
\end{proof}

\section{Embedding a Boolean test in the formula}\label{sec:embedding}
To make the free identifier bits evaluate independent copies of a Boolean test, we use fixed helper clauses to force selected variables false.
The remaining positive clauses can then be satisfied exactly when each contains a variable outside the forced set.
For $a\in\bits^r$, we also write $a$ for its unsigned integer value in the bit order of Definition~\ref{def:lit}.

\begin{lemma}[Sparse-helper restriction for an arbitrary block]\label{lem:embedg}
Let $1\le r\le w-1$ and let $g:\bits^r\to\bits$ be any Boolean function.
Put $M=m-|g^{-1}(0)|$ and assume $M\ge1$.
There is a restriction fixing exactly $n-3Mr$ bits of $F_w$ whose cofactor is
\[
G=\bigwedge_{c=1}^{M}\bigl(g(x_{c,1})\vee g(x_{c,2})\vee g(x_{c,3})\bigr).
\]
Here each $x_{c,j}\in\bits^r$ is a separate block of free input bits, and their total number is $L=3Mr$.
\end{lemma}
\begin{proof}
Call the identifier $v=2^{w-1}+a$ forbidden when $g(a)=0$.
For each forbidden $v$, insert the fixed clause
$(\neg X_v\vee\neg X_v\vee\neg X_v)$.
This forces $X_v=0$, using one record for each of the $|g^{-1}(0)|$ forbidden identifiers.

Use the remaining $M$ records as positive clauses by setting every sign bit to zero.
In each field, set the highest identifier bit to one and positions $r,\ldots,w-2$ to zero, leaving the lowest $r$ bits free.
The range $r\le w-1$ makes these choices consistent.
There are exactly $3Mr$ free bits, so exactly $n-3Mr$ bits have been fixed.

Every free identifier lies between $2^{w-1}$ and $2^{w-1}+2^r-1\le m$, so none is zero or out of range.
If a positive clause has three forbidden identifiers, the fixed clauses force it false.
If every positive clause contains an identifier that is not forbidden, setting every unforced variable true and every forbidden variable false gives a single assignment satisfying every clause.
In a free field, an identifier is not forbidden exactly when its low bits make $g$ equal to one.
The satisfiability condition is therefore the displayed function $G$.
\end{proof}

\paragraph{A small example.}
Take $w=5$, $r=2$ and $g(a_0,a_1)=a_0\vee a_1$.
Of the four possible identifiers $16,17,18,19$, only $16$ is forbidden, so one clause forces $X_{16}=0$.
The other $30$ clauses use positive literals with two free identifier bits each.
They are jointly satisfiable exactly when each clause contains an identifier other than $16$.
Section~\ref{sec:blocks} constructs tests that retain their input dependence after restrictions, as required for the lower bound.

To count the gates removed by fixing these bits, we first show that every description bit can affect the answer.
\begin{lemma}[Essentiality of every encoding bit]\label{lem:essential}
Every bit in Definition~\ref{def:enc} is essential to $F_w$.
\end{lemma}
\begin{proof}
Select any clause record and any of its fields.
Fill every otherwise unused record with the tautology $(X_1\vee\neg X_1\vee X_1)$.
The assumption $w\ge5$ gives $m\ge31$, so the additional records below are available.

For a sign bit, let the selected literal initially be $X_2$ and its two companions be $X_1$.
In two other records, force $X_1=X_2=0$ using negative repeated-literal clauses.
The selected clause is then false, so the formula is unsatisfiable.
Flipping only the selected sign changes $X_2$ to $\neg X_2$ and makes the formula satisfiable.

For identifier bit position $j\in\{0,\ldots,w-1\}$, set all three literals of the selected record initially to positive $X_m$.
In another record, force $X_m=0$, making the formula unsatisfiable.
All bits of $m=2^w-1$ are one, so flipping the selected bit gives identifier $m-2^j$, which lies in $\{1,\ldots,m-1\}$.
Setting $X_m=0$ and $X_{m-2^j}=1$ now satisfies the formula.
Each witness changes only the selected encoding bit, proving essentiality.
\end{proof}
Lemmas~\ref{lem:floor} and~\ref{lem:essential} give the lower bound $n-1$.
We next count the gates removed by fixing essential inputs and those required by the remaining function.

\section{Transferring a lower bound through a restriction}\label{sec:transfer}
We first normalize a circuit by simplifying its gates and wiring without increasing its size, then bound the number of gates removed when essential inputs are fixed.
\begin{definition}[Gate types]\label{def:type}
An essential binary truth table depends on both formal arguments.
There are two parity tables, $x\oplus y\oplus c$, where $\oplus$ is exclusive OR and $c\in\bits$.
The other eight essential tables are nonlinear and have the form
$((x\oplus a)\wedge(y\oplus b))\oplus c$ for $a,b,c\in\bits$.
A degenerate table is constant or depends on only one argument.
Write $X(C)$ and $A(C)$ for the numbers of parity and nonlinear gates in $C$.
\end{definition}
The eight nonlinear tables output one on either one or three rows; the classification concerns the gate's truth table even when the actual predecessor signals are related.

A nonlinear gate has an argument value that makes its other argument irrelevant.
\begin{definition}[Blocking value]\label{def:blocking}
A blocking value at an operand of a nonlinear gate makes the output constant for both values of the other operand.
In the nonlinear form of Definition~\ref{def:type}, the blocking values are $a$ and $b$, and the constant output is $c$.
\end{definition}
Fixing an operand to its blocking value can therefore remove the gate and propagate a constant to its successors.

\begin{definition}[Normalized circuit]\label{def:norm}
A circuit is normalized if every gate has an essential binary table, its two predecessors are distinct, and every gate reaches the output.
It must also have no input triangle.
An input triangle consists of gates $P=p(x,u)$ and $Q=q(x,P)$, where $x$ is a primary input and $u$ is another predecessor.
\end{definition}
Removing an input triangle lets $Q$ compute directly from $(x,u)$ using a single binary truth table.
These simplifications are customary in gate elimination~\cite{Zwi91,GKST16}; we give the details needed here.

\begin{lemma}[Normalization]\label{lem:normalize}
If the output function has at least two essential inputs, it has a normalized circuit of no greater size than the original circuit.
\end{lemma}
\begin{proof}
Here a literal of a signal means that signal or its complement.
\begin{proofsteps}
\item \emph{Remove gates that compute no essential binary operation.}
Delete gates with no path to the output.
A constant operand, repeated operand or degenerate table leaves a constant or a literal of one predecessor, so substitute that constant or bypass that gate.
In the latter case, incorporate any negation into the truth tables of the gates receiving that signal.
Since the output has at least two essential inputs, its surviving signal is a literal of a binary gate.
Choose that gate as output, prune its successors and absorb any final negation into its table.
\item \emph{Replace an input triangle.}
For $P=p(x,u)$ and $Q=q(x,P)$, replace the table at $Q$ by $\psi(x,u)=q(x,p(x,u))$.
Retain $P$ if other gates use its output.
If $q$ is nonlinear, blocking it at $x$ gives a constant cofactor of $\psi$.
If $q$ is parity and $p$ is nonlinear, blocking $p$ at $x$ gives a constant cofactor.
If both are parity, the two occurrences of $x$ cancel.
Thus $\psi$ is nonlinear or degenerate; it is never an essential parity table.
Apply Step~1 again.
\item \emph{Check termination.}
All deletions and bypasses preserve the original predecessor ordering, and no step increases the number of gates.
Between deletions, a triangle replacement changes the predecessor of $Q$ from $P$ to the earlier vertex $u$.
This strictly decreases the sum of predecessor indices over all wires, so there are only finitely many replacements and deletions.
\end{proofsteps}
At termination the circuit is normalized, the function is unchanged and no gate has been added.
\end{proof}

We can therefore normalize a circuit of minimum size before applying a restriction.
\begin{lemma}[Gates removed by a restriction]\label{lem:closure}
In a normalized circuit, fix $t$ inputs that were essential before restriction.
Suppose the cofactor has at least two essential inputs.
Simplification removes at least $t$ gates.
\end{lemma}
\begin{proof}
Let $d$ count wires leaving the fixed inputs, whose essentiality gives $d\ge t$.
Let $S$ contain those inputs and all $h$ gates made constant by propagating constants.
The output is outside $S$, because the cofactor still has two essential inputs.

Let $e$ count wires with both endpoints in $S$, and let $b$ count wires leaving $S$.
Each constant gate receives at most two wires from $S$, so $e\le2h$.
Every gate in $S$ has an outgoing wire, since the original circuit is normalized and its output is outside $S$.
Thus $e+b\ge d+h$, and hence $h+b\ge d$.

The $b$ wires leaving $S$ enter distinct gates: if two entered one gate, both its operands would be constant and it would belong to $S$.
Each receiving gate has a constant operand and can be bypassed or replaced by a constant using Lemma~\ref{lem:normalize}.
These $b$ gates are distinct from the $h$ constant gates, so at least $h+b\ge d\ge t$ gates are removed.
\end{proof}
Applying this count to the SAT restriction gives the following transfer inequality.
\begin{lemma}[Transfer]\label{lem:transfer}
If a restriction fixing $n-L$ bits of $F_w$ yields a function $G$ depending on all $L\ge2$ free bits, then
$\C(F_w)\ge n-L+\C(G)$.
\end{lemma}
\begin{proof}
Normalize a minimum circuit using Lemma~\ref{lem:normalize}.
Every fixed encoding bit is essential before restriction by Lemma~\ref{lem:essential}, so Lemma~\ref{lem:closure} removes at least $n-L$ gates.
The remaining circuit computes $G$, so it has at least $\C(G)$ gates; adding the removed gates gives the inequality.
Both costs are measured within the same original circuit, so no disjointness assumption about its computations is needed.
\end{proof}

\section{Boolean tests that survive restrictions}\label{sec:blocks}
The transfer lemma reduces the SAT problem to a circuit bound for the free identifier fields, so we choose each field's Boolean test to support two parts of the gate count.
High algebraic degree forces many nonlinear gates, while distinct cofactors keep input settings distinguishable as gate elimination fixes other bits.

\begin{definition}[Algebraic degree]\label{def:deg}
Represent a Boolean function by its unique multilinear polynomial over the two-element field $\mathbb F_2$.
Here addition is XOR, multiplication is AND, and multilinear means that each variable has exponent at most one.
The algebraic degree $\degF f$ is the largest number of variables in a monomial with nonzero coefficient.
We assign degree zero to constants.
An $r$-variable function has full degree if its degree is $r$~\cite[Section~2.2.1]{Carlet}.
\end{definition}
Counting accepting assignments modulo two detects full degree without constructing the polynomial.
\begin{lemma}[Test for full degree]\label{lem:odd}
For $r\ge1$, an $r$-variable function has full degree exactly when it has an odd number of accepting assignments.
\end{lemma}
\begin{proof}
Sum the polynomial values over the Boolean cube.
A monomial on $j$ variables is one on exactly $2^{r-j}$ assignments; if $j<r$, that number is even and its contribution vanishes in $\mathbb F_2$.
For $j=r$, the contribution is exactly the coefficient of the product of all $r$ variables.
The sum counts accepting assignments modulo two, so their number is odd exactly when the full-degree coefficient is one.
\end{proof}
Exactly half of all Boolean truth tables have full degree.

The SAT embedding combines tests on disjoint identifier fields by AND and OR.
\begin{definition}[Read-once composition]\label{def:tree}
A block is a Boolean function on its own set of input bits.
A read-once AND/OR composition is a rooted binary tree of AND and OR operations.
Its leaves are blocks on pairwise disjoint sets of inputs, and each block occurs once.
Write $B$ for the number of leaves and $L$ for the total number of input bits.
\end{definition}
Here read-once describes the tree defining the function; circuits computing that function may still share gates arbitrarily.

Schnorr's bound relates polynomial degree to the number of nonlinear gates~\cite{Sch89}; we use its binary-circuit formulation due to Kojevnikov and Kulikov~\cite[Lemma~1]{KK10}.
\begin{lemma}[Degree requires nonlinear gates]\label{lem:nonlin}
If $f$ has algebraic degree $D\ge1$, every $B_2$ circuit computing $f$ contains at least $D-1$ nonlinear gates.
\end{lemma}
\begin{proof}
List the nonlinear gates as $z_1,\ldots,z_A$ in predecessor order.
Regard their outputs as independent formal Boolean variables.
Every other gate is affine over $\mathbb F_2$, meaning a constant plus a sum of individual variables.
Thus every signal before $z_j$ is affine in the original inputs and $z_1,\ldots,z_{j-1}$.
The polynomial $p_j$ defining $z_j$ has total degree at most two in those symbols.
The final output is affine in the original inputs and all the $z_j$.

Work with Boolean polynomials, replacing $u^2$ by $u$ for every current variable $u$, including each formal gate variable.
This makes the polynomial multilinear after every substitution.
If its current degree is at most $d\ge1$, separate the terms containing $z_j$:
\[
P=z_jU+V,\qquad
z_j\notin\operatorname{Var}(U)\cup\operatorname{Var}(V),\qquad
\deg U\le d-1,\quad \deg V\le d.
\]
Here $\operatorname{Var}(U)$ is the set of variables appearing in $U$.
Replacing $z_j$ by $p_j$ gives $p_jU+V$, of degree at most $d+1$.
Making the result multilinear cannot increase its degree.

Substitute $z_A,z_{A-1},\ldots,z_1$ in that order.
Each $p_j$ uses only earlier gate symbols, so the substitutions leave only original inputs.
Starting at degree at most one, the $A$ substitutions leave degree at most $A+1$.
The resulting polynomial represents $f$, so $D\le A+1$, equivalently $A\ge D-1$.
\end{proof}
This bound remains valid with arbitrary sharing; the reverse substitution counts each nonlinear gate once.

Disjointness makes the degrees of the blocks add when they are combined by AND or OR.
\begin{proposition}[Degree of a read-once composition]\label{cor:degL}
A composition of functions of degrees $d_1,\ldots,d_B\ge1$ on disjoint inputs has degree $d_1+\cdots+d_B$.
In particular, a composition of full-degree blocks on $L$ inputs has degree $L$.
Every circuit for that composition has at least $L-1$ nonlinear gates.
\end{proposition}
\begin{proof}
Let $f$ and $g$ use disjoint variables and have degrees $d,e\ge1$.
A product of a degree-$d$ monomial of $f$ with a degree-$e$ monomial of $g$ has degree $d+e$.
Disjoint variable sets uniquely determine the two factors of that monomial, so these terms cannot cancel.
Thus $fg$ has degree $d+e$.
The polynomial for OR is $f+g+fg$.
Since $d+e$ exceeds both $d$ and $e$, its highest-degree part is again that of $fg$.

At every parent, the two child subtrees still use disjoint input sets, so their degrees add.
Induction gives $d_1+\cdots+d_B$ at the root.
For full-degree blocks the sum is $L$, so Lemma~\ref{lem:nonlin} gives at least $L-1$ nonlinear gates in any circuit for the composition.
\end{proof}
Gate elimination also needs input dependence to persist after restrictions.
\begin{definition}[Cofactor conditions]\label{def:cond}
For a Boolean function, use the following three conditions.
\begin{itemize}[leftmargin=*,itemsep=2pt]
\item (E1): after fixing any one bit, every remaining bit is essential.
\item (E2): after fixing any set of at most two bits, every remaining bit is essential.
\item (S4): for any two distinct bits, the four cofactors obtained by fixing them are pairwise distinct.
\end{itemize}
\end{definition}
We use (E2) before a one-bit restriction and (E1) afterward.

\begin{definition}[Robust block]\label{def:robust}
For integers $3\le k\le r$, a block on $r$ inputs is $k$-robust if every restriction leaving at least $k$ inputs satisfies (S4).
\end{definition}
This is strong two-dependence in the terminology of Lachish and Raz~\cite[Definitions~3.1--3.2]{LR01}.
It suffices to check restrictions leaving exactly $k$ inputs, because equality of two cofactors persists under every common further restriction.
A failure on more than $k$ inputs therefore gives a failure on exactly $k$: keep the selected pair free and fix other inputs until exactly $k$ remain, as $k\ge3$ permits.

\begin{lemma}[Cofactors across blocks]\label{lem:separate}
Suppose every leaf of a read-once composition is a restriction of a $k$-robust block.
If every leaf retains at least $k+2$ bits, the composition has (E1), (E2) and (S4).
If every leaf retains at least $k+1$ bits, the composition still has (E1).
\end{lemma}
\begin{proof}
\begin{proofsteps}
\item \emph{Show that every relevant leaf can take both values.}
A robust leaf with at least $k$ remaining bits is essential in each bit.
Otherwise a pair containing an inessential bit would have two equal cofactors, contrary to (S4).
Each leaf is therefore nonconstant.
Every subtree is also nonconstant: make both its children zero or both one to obtain the desired parent value.
Their disjoint inputs allow these choices simultaneously.
\item \emph{Expose a leaf and prove the essentiality assertions.}
On the path from the selected leaf to the root, set every sibling to one at an AND parent or zero at an OR parent.
The root then equals the selected leaf.
This remains possible after restrictions leaving at least $k$ bits in every leaf.
Starting with $k+2$ bits permits any two bits to be fixed, proving (E2) and hence (E1).
Starting with $k+1$ permits one bit to be fixed, proving the final assertion.
\item \emph{Separate four cofactors for bits in one leaf.}
Expose that leaf using Step~2.
Its four cofactors are distinct by robustness.
If two cofactors of the full composition were equal, exposing the leaf would preserve that equality, a contradiction.
\item \emph{Separate four cofactors for bits in two leaves.}
Set every subtree branching off their two paths to the neutral value from Step~2.
The surviving operation is AND or OR at the paths' first common ancestor.
Each selected bit gives two distinct cofactors of its leaf, by essentiality.
Each of those cofactors is nonconstant, by essentiality after that bit is fixed.
For AND, accepting assignments are pairs drawn from the two nonempty accepting sets.
Each set is recovered by taking the corresponding coordinate of those pairs.
The two choices for each set therefore give four distinct sets of accepting assignments.
For OR, apply the same reasoning to the nonempty rejecting sets~\cite[Section~1.2]{KN}.
\end{proofsteps}
\end{proof}

\subsection{Constructing full-degree robust blocks}
A counting argument gives full-degree tests whose robustness threshold $k$ grows logarithmically with the block length; the construction using a weighted sum modulo a prime gives a threshold of square-root order.
For the SAT restriction, it suffices to prove that such tests exist.
The counting proof uses random truth tables and a union bound to separate cofactors, as in Kabanets~\cite[Section~3]{Kab00}, and also requires full degree.
\begin{proposition}[Counting construction]\label{prop:random}
Let $r\ge k+2$ and $k\ge3$.
If $2^{k-2}>\log_2(6r(r-1))+r\log_2 3$, there is a $k$-robust function on $r$ inputs of full $\mathbb F_2$-degree $r$.
\end{proposition}
\begin{proof}
Choose $g$ uniformly from all truth tables on $r$ bits.
After a restriction with $j\ge k$ free bits, the four settings of a selected pair leave four disjoint sets of assignments.
The corresponding truth tables are independent uniform functions on $j-2$ bits.
There are $\binom42=6$ pairs of these functions.
Each pair coincides with probability $2^{-2^{j-2}}$, so any coincidence has probability at most $6\cdot2^{-2^{k-2}}$.

There are at most $3^r$ restrictions, since each bit is free, fixed to zero, or fixed to one, and at most $\binom r2$ selected pairs for each restriction.
The union bound therefore gives failure probability at most
\[
3^r\cdot3r(r-1)\cdot2^{-2^{k-2}}<\frac12.
\]
The strict inequality is exactly the stated logarithmic hypothesis after exponentiating base two.

Exactly half the truth tables have an odd number of ones: toggling any one entry pairs odd and even tables.
By Lemma~\ref{lem:odd}, full degree thus has probability $1/2$.
The failure bound therefore leaves a positive probability of full degree and robustness together.
\end{proof}
For large $r$, taking $k=\log_2 r+O(1)$ makes the inequality hold.

Weighted sums modulo a prime, combined with subset-sum coverage, were used by Savick\'y and \v{Z}\'ak to distinguish cofactors after restrictions~\cite[Theorem~2.5 and Section~3]{SZ96}.
Here we test equality to a single residue, choosing the modulus to separate all four pair settings and the residue to give full degree.
\begin{proposition}[Weighted modular blocks]\label{prop:explicit}
Let $r\ge4$ and let $p$ be the least prime greater than $2r$.
Let $m_p$ be the least nonnegative integer satisfying $1+m_p(m_p+1)/2\ge p$.
Put $k=m_p+2\le\lceil\sqrt{8r}\rceil+2$.
For $r\ge k+2$ and every residue $t$ modulo $p$, the function
\[
g_{r,t}(a_1,\ldots,a_r)=\Bigl[\textstyle\sum_{i=1}^r i\,a_i\equiv t\pmod p\Bigr]
\]
is $k$-robust.
For at least one residue $t$, it has full degree $r$.
Here $[P]$ equals one if the statement $P$ holds and zero otherwise.
\end{proposition}
The weights $1,\ldots,r$ distinguish the four settings of any two bits, while sufficiently many remaining weights guarantee that each corresponding residue is attainable.
\begin{proof}
\begin{proofsteps}
\item \emph{Identify the four target residues.}
After fixing some bits, let $F$ be the free index set and let $\tau$ be $t$ minus the contribution of the fixed bits, modulo $p$.
Assume $|F|\ge k$.
For distinct $i,j\in F$, the four cofactors have the form $[S\equiv\tau-\sigma\pmod p]$.
Here $\sigma\in\{0,i,j,i+j\}$ and $S$ is the weighted sum over $F\setminus\{i,j\}$.
The four targets are distinct: $i,j$ are distinct positive integers and $i+j\le2r<p$.
\item \emph{Show that all four targets occur.}
Let $H=F\setminus\{i,j\}$, viewed as a set of residues modulo $p$.
Then $|H|\ge m_p$ and $H\cap(-H)=\varnothing$, because all weights lie in $[1,r]$ and $2r<p$.
Balandraud's subset-sum theorem states that, for an odd prime $p$ and such a set $H$,
\[
|\Sigma(H)|\ge \min\{p,\,1+|H|(|H|+1)/2\}.
\]
Here $\Sigma(H)$ contains the sums of all subsets of $H$, including the empty subset~\cite[Theorem~3]{Bal17}.
The definition of $m_p$ makes the right side $p$, so each target residue is attained by an assignment to the remaining free bits.
Each cofactor is one on a nonempty set of assignments, and these four sets are pairwise disjoint, so the cofactors are distinct.
This proves robustness.
\item \emph{Bound the threshold.}
Bertrand's postulate gives a prime strictly between $2r$ and $4r$~\cite{Che52}.
Since $p$ is the least prime above $2r$, we have $p<4r$.
Setting $q=\lceil\sqrt{8r}\rceil$ gives $q(q+1)/2>4r>p-1$, so $m_p\le q$ and $k\le q+2$.
\item \emph{Find an odd residue count.}
Let $N_t$ count assignments with $\sum_i i a_i\equiv t\pmod p$.
In $\mathbb F_2[z]/(z^p-1)$, the parity of these counts is recorded by
\[
\sum_{t=0}^{p-1}(N_t\bmod2)z^t
\equiv \prod_{i=1}^r(1+z^i)\pmod{z^p-1}.
\]
Each factor chooses whether to include weight $i$, and reduction modulo $z^p-1$ collects sums with the same residue.
Write $\Phi_p(z)=1+z+\cdots+z^{p-1}$, so $z^p-1=(z+1)\Phi_p(z)$ over $\mathbb F_2$.
Since $p$ is odd, the derivative of $z^p-1$ in characteristic two is $z^{p-1}$.
It is coprime to $z^p-1$, so the latter polynomial has no repeated factors.
Work in a splitting field of $z^p-1$ over $\mathbb F_2$, where the polynomial factors into linear terms.
Since $\Phi_p(1)=1$, its roots are $p$-th roots of unity other than one.
Each has multiplicative order $p$, because $p$ is prime.
Such a root cannot satisfy $1+z^i=0$ for $1\le i\le r<p$.
Thus no irreducible factor of $\Phi_p$ divides any factor of the product.
In particular, $\Phi_p$ does not divide the product, whose remainder modulo $z^p-1$ is nonzero.
At least one $N_t$ is therefore odd, and Lemma~\ref{lem:odd} gives full degree for that $g_{r,t}$.
\item \emph{Specify the residue effectively.}
Start with the length-$p$ parity vector having a one at residue zero and zeros elsewhere.
For $i=1,\ldots,r$, add its cyclic shift by $i$ to itself in $\mathbb F_2^p$.
Equivalently, update the polynomial vector by $v\leftarrow v+z^iv$ modulo $z^p-1$.
After $r$ updates its entries are the parities of all $N_t$, so choose the least residue with a one.
Once $p$ is fixed, this recurrence costs $O(rp)=O(r^2)$ bit operations.
\end{proofsteps}
\end{proof}
This quadratic bound assumes that the prime is supplied; Section~\ref{sec:formalization} describes the formalized cost of finding the prime as well as constructing the block.

\section{Counting gates under a restriction}\label{sec:accounting}
We measure the effect of a restriction by a weighted gate count, subtracting one unit for each of two kinds of credit attached to primary inputs.
An input earns the first credit when its only successor is nonlinear, and the second when that gate's only successor is also nonlinear.
To track how these credits change under restriction, we record which gates use each signal during simplification.

We first distinguish the number of successor gates from the number of operand positions they use.
\begin{definition}[Successor set and fan-out]\label{def:succ}
For a vertex $v$ in Definition~\ref{def:C}, let $\succS(v)$ be the set of gates using $v$ as an operand.
Its \emph{fan-out} is $d(v)=|\succS(v)|$.
For a set $Y$ of vertices, put $\succS(Y)=\bigcup_{v\in Y}\succS(v)$.
These sets refer to the current circuit.
They include any gates temporarily present that have no path to the output.
\end{definition}
A gate using a signal twice counts as one successor, although each gate in the circuits simplified below has two distinct predecessors.

\begin{definition}[Resolution under a restriction]\label{def:resolve}
Let $C$ compute $f$.
Assume that its gates have distinct predecessors and that its output gate has no successors.
Intermediate gates may have degenerate tables, as in Definition~\ref{def:type}.
Let $\rho$ fix the input set $I$, and suppose that $f|_\rho$ has at least two essential inputs.

Process nodes in topological order, so every predecessor is processed first.
Assign each node $v$ a \emph{resolved signal} $\sigma(v)$.
This is a constant or a literal of a node, meaning that node's value or its complement.
A fixed input resolves to its assigned constant; a free input $z$ resolves to $z$.
For $G=g(a,b)$, substitute the resolved operands into $g$.
Test the resulting table with distinct underlying node names as independent formal Boolean arguments, regardless of relations between the signals they compute.
The words \emph{becomes constant} and \emph{depends} in the following rules refer to this local table test.
\begin{itemize}
\item If $g(\sigma(a),\sigma(b))$ is constant, $G$ \emph{becomes constant} and $\sigma(G)$ is that constant.
\item If it depends on exactly one underlying node $N$, $\sigma(G)$ is the corresponding literal of $N$.
We say that $G$ is \emph{bypassed} and \emph{passes} $N$.
\item If it depends on two distinct underlying nodes, $G$ \emph{survives resolution} and $\sigma(G)=G$.
Replace its operands by those nodes and absorb any complements into its table.
\end{itemize}
\end{definition}
The rule covers constant operands, degenerate tables, and a \emph{collapse}, where both operands resolve to literals of the same node and the resulting table gives either a constant or a literal.

Resolution determines the output signal; pruning then removes gates that no longer contribute to it.
\begin{definition}[Simplified circuit]\label{def:simp}
Under Definition~\ref{def:resolve}, suppose that $f|_\rho$ depends on at least two inputs.
The original output resolves to a literal of a gate $N$ that survives resolution.
Designate $N$ as the new output.
If the literal is complemented, complement the table at $N$.
Prune every gate without a path to $N$.
In particular, no successor of $N$ remains, so changing its table affects no other retained gate.

Call the resulting circuit $C^\rho$ the \emph{simplified circuit}.
An original gate is \emph{eliminated} if it is absent from $C^\rho$.
This includes gates that become constant, are bypassed, or survive resolution but are pruned.
Write $N(C^\rho)$ for a normalization of $C^\rho$ by Lemma~\ref{lem:normalize}.
\end{definition}
For the rest of this section, circuits retain the distinct-predecessor and output conventions of Definition~\ref{def:resolve}, and every application of resolution assumes that the cofactor has at least two essential inputs.

\begin{lemma}[Resolution and dependence]\label{lem:disconnect}
The simplified circuit of Definition~\ref{def:simp} computes $f|_\rho$ and preserves the type of every retained gate.
If every path from a free input to the output meets a gate constant under $\rho$, then $f|_\rho$ is independent of that input.
\end{lemma}
The path criterion holds for any circuit and restriction, without the two-essential-input hypothesis.
It also applies when a gate computes a constant function of the free inputs but the local resolution test does not detect this.
\begin{proof}
Each resolved operand has exactly its original value under $\rho$, so induction in topological order preserves every gate value and the output.
Subsequent pruning removes only gates with no path to that output.

A retained gate has two distinct resolved operands, and its table changes only by input complements or by complementing the designated output.
Both operations preserve the parity and nonlinear types of Definition~\ref{def:type}.

Replace each gate constant under $\rho$ by its value.
If every path from a free input meets such a gate, no remaining path carries that input to the output.
The cofactor is therefore independent of it.
\end{proof}

The following lemma tracks a signal through the gates that pass it.
\begin{lemma}[Passing]\label{lem:pass}
Use Definitions~\ref{def:succ}--\ref{def:simp}.
Let $N$ be a free input or a gate that survives resolution.
Let $S_N$ be the set of nodes whose resolved signal is a literal of $N$; in particular, $N\in S_N$.
\begin{enumerate}
\item[(a)] Every node of $S_N\setminus\{N\}$ is a gate with an operand in $S_N$.
\item[(b)] The successors of $N$ in $C^\rho$ are exactly the gates of $C^\rho$ with an original operand in $S_N$.
\item[(c)] If $N\in Y\subseteq S_N$ and no gate of $\succS(Y)\setminus Y$ lies in $S_N$, then $S_N=Y$.
\end{enumerate}
\end{lemma}
Part (c) lets us prove that a collection contains every node passing $N$ by checking that no edge leaving it leads to another such node.
\begin{proof}
For (a), a node of $S_N\setminus\{N\}$ cannot be a primary input.
Write it as $G=g(a,b)$.
For $N$ to occur in the substituted truth table $g(\sigma(a),\sigma(b))$, at least one resolved operand must be a literal of $N$.

For (b), each operand of a gate retained in $C^\rho$ is the underlying node of an original operand's resolved signal.
It is $N$ exactly when that original operand lies in $S_N$.

For (c), if $S_N\setminus Y$ were nonempty, let $G$ be its first node in topological order.
By (a), $G$ has a predecessor in $S_N$.
That predecessor occurs earlier, so it lies in $Y$.
Thus $G\in\succS(Y)\setminus Y$, contrary to the hypothesis.
\end{proof}

\begin{definition}[First credit]\label{def:D}
A primary input $z$ earns one credit if $\succS(z)$ consists of a single nonlinear gate $W_z$.
Call $z$ a \emph{D-variable} and $W_z$ its \emph{D-gate}.
Write $\mathcal D(C)$ for the set of D-variables and $D(C)=|\mathcal D(C)|$ for its cardinality.
\end{definition}
The credit depends on the current successor set, so it can appear or disappear after a restriction.

\begin{samepage}
The dependence conditions prevent fixing one or two other inputs from cutting off a credited input's effect on the output.
\begin{lemma}[Forbidden configurations]\label{lem:forbid}
Let $C$ compute $f$, with credits as in Definition~\ref{def:D} and conditions as in Definition~\ref{def:cond}.
\begin{enumerate}
\item[(a)] Under (E1), no gate with two primary operands is a D-gate.
Consequently, distinct D-variables have distinct D-gates and $D(C)\le A(C)$.
\item[(b)] Under (E2), no successor of an essential gate $T$ with two primary operands is a D-gate.
\end{enumerate}
\end{lemma}
\end{samepage}
These exclusions will bound how many lost credits can be assigned to one eliminated gate.
\begin{proof}
For (a), suppose $W=W_z$ has operands $z$ and a distinct primary input $v$.
Fixing $v$ to its blocking value for $W$ makes the gate constant.
Every path from $z$ begins with $z\to W$, so Lemma~\ref{lem:disconnect} makes the cofactor independent of $z$.
This contradicts (E1).
If two D-variables shared a D-gate, its operands would be those two primary inputs.
Thus $z\mapsto W_z$ is injective.
Since every $W_z$ is nonlinear, $D(C)\le A(C)$.

For (b), suppose $T$ has primary operands $x,y$, and its successor $W$ is the D-gate of $z$.
The operands of $W$ are $T,z$.
Also $z\notin\{x,y\}$: its only successor is $W$, whereas both $x$ and $y$ feed $T$.
Since the essential table at $T$ takes both values, choose $(x,y)$ so that $T$ blocks $W$.
Every path from $z$ then meets the constant gate $W$, so Lemma~\ref{lem:disconnect} contradicts (E2).
\end{proof}

The second credit records one more nonlinear gate on the same single-successor chain.
\begin{definition}[Second credit]\label{def:star}
An input $z\in\mathcal D(C)$ earns a second credit if $\succS(W_z)=\{W_z^2\}$ and $W_z^2$ is nonlinear.
Call $z$ a \emph{$D^*$-variable}.
Write $\mathcal D^*(C)$ for their set and $D^*(C)$ for its cardinality.
A gate has the \emph{first role} for $z$ if it is $W_z$, and the \emph{second role} if it is $W_z^2$.
\end{definition}
The superscript in $W_z^2$ names the second gate rather than a square, and a $D^*$-variable earns both credits.

\begin{definition}[Two-level potential]\label{def:psi}
Using the gate counts of Definition~\ref{def:type} and the credits of Definitions~\ref{def:D} and~\ref{def:star}, put
\begin{equation}\label{eq:psi}
\Psi_2(C)=7X(C)+4A(C)-D(C)-D^*(C).
\end{equation}
\end{definition}
Eliminating gates or creating credits lowers $\Psi_2$; lost credits offset part of this decrease.

\begin{lemma}[Roles]\label{lem:roles}
Let $C$ compute $f$, with roles as in Definition~\ref{def:star}.
\begin{enumerate}
\item[(a)] Under (E1), a gate with a primary operand cannot have a second role.
A direct successor of a primary input $x$ with $d(x)\ge2$ has neither role.
\item[(b)] Under (E2), a successor of an essential gate with two primary operands has neither role.
\end{enumerate}
\end{lemma}
We call a gate \emph{role-free} if it has neither credit role; such a gate cannot be assigned a lost credit in the accounting below.
\begin{proof}
For (a), suppose $G=g(x,W_z)=W_z^2$ has primary operand $x$.
The input $z$ differs from $x$, since $z$ feeds only $W_z$ whereas $x$ feeds $G$.
Fix $x$ to its blocking value for $G$.
Every path from $z$ passes through $W_z$ and then the now constant $G$.
The cofactor is independent of $z$, contradicting (E1).

Let $G$ be a successor of $x$ with $d(x)\ge2$.
The preceding argument excludes a second role.
If $G=W_z$, then $z\ne x$ because $z$ has only one successor.
The operands of $G$ would be the primary inputs $x,z$, contradicting Lemma~\ref{lem:forbid}(a).

For (b), let $T$ have primary operands $x,y$ and let $S$ be its successor.
Lemma~\ref{lem:forbid}(b) excludes a first role.
If $S=W_z^2$, its operands are $T,W_z$.
Lemma~\ref{lem:forbid}(a) gives $W_z\ne T$.
Also $z\notin\{x,y\}$, because $z$ feeds only $W_z$.
Choose $(x,y)$ so that $T$ blocks $S$.
All paths from $z$ then meet the constant gate $S$, contradicting (E2).
\end{proof}

We account for each lost credit by subtracting it from the contribution of an eliminated gate, calling the total assigned to that gate its \emph{charge}.
\begin{lemma}[Charging for $\Psi_2$]\label{lem:charge2}
Use the potential of Definition~\ref{def:psi} and resolution of Definition~\ref{def:resolve}.
Let $C$ compute a function $f$ satisfying (E1).
Let $\rho$ fix the input set $I$.
Assume that $f|_\rho$ depends on every free input and that at least two inputs remain.

Write $[P]=1$ when a statement $P$ holds, and $[P]=0$ otherwise.
The old credit of input $z$ is
$\lambda(z)=[z\in\mathcal D(C)]+[z\in\mathcal D^*(C)]$.
For $z\notin I$, let $\lambda'(z)$ be its corresponding credit in $C^\rho$.
For $z\in I$, put $\lambda'(z)=0$.
The total gained credit is $\Gamma=\sum_z\max(\lambda'(z)-\lambda(z),0)$.

Give an original gate $G$ weight $w_2(G)$ equal to $7$, $4$, or $0$ according as it is parity, nonlinear, or degenerate.
There are nonnegative integer charges $\chi_2(G)$ on eliminated gates such that
\begin{equation*}\tag{1a}\label{eq:charge}
\Psi_2(C)-\Psi_2(C^\rho)
 =\sum_{G\text{ eliminated}}\bigl(w_2(G)-\chi_2(G)\bigr)+\Gamma.
\end{equation*}
Only nonlinear gates with a first or second role are charged, and each charge is at most two.
If $I\cap\mathcal D(C)=\varnothing$, each charge is at most one and no charged gate becomes constant.
\end{lemma}
Equation~\eqref{eq:charge} separates the decrease into gate weights, lost credits, and gained credits.
If no D-variable is fixed, an eliminated nonlinear gate contributes at least three units.
A role-free nonlinear gate, or one that becomes constant, contributes all four.
\begin{proof}
Allocate each input's lost credit to eliminated gates on its original credit chain, then bound each gate's charge.
\begin{proofsteps}
\item \emph{Separate credit losses from gains.}
Let $\Lambda=\sum_z\max(\lambda(z)-\lambda'(z),0)$ be the total lost credit.
The original value of $D+D^*$ minus its value after restriction is $\Lambda-\Gamma$.
By Lemma~\ref{lem:disconnect}, the decrease in $7X+4A$ is the sum of $w_2(G)$ over eliminated gates.
It suffices to allocate exactly $\Lambda$ units to those gates.
Any input with $\lambda(z)>\lambda'(z)$ belongs to $\mathcal D(C)$.

\item \emph{Allocate a loss when $z$ is free and $W_z$ remains.}
Since $W_z$ survives resolution, $\sigma(W_z)=W_z$, and Lemma~\ref{lem:pass}(c), with $Y=\{z\}$, gives $S_z=\{z\}$.
Part (b) makes $W_z$ the only successor of $z$ in $C^\rho$; it remains nonlinear, so the first credit persists.
The loss is therefore one, from $z$ losing its second credit.

The gate $W_z^2$ must be eliminated: if it remained, Lemma~\ref{lem:pass}(c) would give $S_{W_z}=\{W_z\}$.
Part (b) would then make the nonlinear gate $W_z^2$ the sole successor of $W_z$ in $C^\rho$, preserving the second credit.
We therefore charge the lost unit to $W_z^2$.

\item \emph{Allocate a loss when $z$ is free and $W_z$ is eliminated.}
If $z$ still has a first credit in $C^\rho$, the positive loss is one; charge it to $W_z$.
Otherwise, $z$ loses all $\lambda(z)$ credits.
Charge $W_z$ once and, when $z\in\mathcal D^*(C)$, charge $W_z^2$ once.

To justify charging $W_z^2$, suppose that it remains.
If $W_z\notin S_z$, Lemma~\ref{lem:pass}(c) gives $S_z=\{z\}$.
Part (b) then leaves $z$ with no successor in $C^\rho$, contradicting essentiality.
Thus $W_z\in S_z$.
Apply part (c) to $Y=\{z,W_z\}$.
Its only outgoing successor is $W_z^2$, which survives resolution and does not pass $z$.
Hence $S_z=\{z,W_z\}$.
By part (b), the only successor of $z$ in $C^\rho$ is the nonlinear gate $W_z^2$.
This would give $z$ a first credit, contrary to assumption.
Therefore $W_z^2$ is eliminated.

\item \emph{Allocate a loss when $z$ is fixed.}
Here $\lambda'(z)=0$, so the loss is $\lambda(z)$.
The gate $W_z$ has a constant operand and is eliminated; charge it once.
If $z$ had a second credit, charge $W_z^2$ when that gate is eliminated.
If $W_z^2$ remains, assign the second unit to $W_z$ instead.

\item \emph{Recover the potential identity.}
Steps 2--4 allocate exactly $\lambda(z)-\lambda'(z)$ units for each input losing credit.
Every allocation goes to an eliminated nonlinear gate with a role for that input.
Define $\chi_2(G)$ as the number of units assigned to $G$.
Then $\sum_G\chi_2(G)=\Lambda$.
Substitution in Step 1 proves equation~\eqref{eq:charge}.

\item \emph{Bound each gate's charge by two.}
If $G=W_z$, it has a primary operand, so Lemma~\ref{lem:roles}(a) excludes every second role at $G$.
Lemma~\ref{lem:forbid}(a) makes $z$ unique.
Only $z$ charges $G$, at most once when free and at most twice when fixed.

Otherwise $G$ can be charged only through second roles.
For each such role, one operand is a first gate $W_z$.
These gates determine distinct inputs by Lemma~\ref{lem:forbid}(a).
A binary gate has two operands, and each associated input assigns at most one unit.
Thus $\chi_2(G)\le2$.

\item \emph{Identify the only possible double charge when no D-variable is fixed.}
Assume $I\cap\mathcal D(C)=\varnothing$.
Step 4 never applies, so a first gate receives at most one unit.
A double charge would require a common second gate
$G=g(W_{z_1},W_{z_2})=W_{z_1}^2=W_{z_2}^2$ for distinct free inputs $z_1,z_2$.
It is enough to show that this gate remains in $C^\rho$.

\item \emph{Show that the common second gate survives resolution.}
For $i=1,2$, the gate $W_{z_i}$ cannot become constant.
Every path from $z_i$ passes through it, and $f|_\rho$ depends on $z_i$.
Nor can $W_{z_i}$ pass a node other than $z_i$.
That would give $S_{z_i}=\{z_i\}$ by Lemma~\ref{lem:pass}(c).
Part (b) would then leave $z_i$ without a retained successor, contrary to its essentiality.
Thus $\sigma(W_{z_i})$ is either $W_{z_i}$ or a literal of $z_i$.

The four nodes $z_1,z_2,W_{z_1},W_{z_2}$ are distinct.
The first two are distinct inputs; the latter two are distinct gates by Lemma~\ref{lem:forbid}(a).
The operands of $G$ therefore resolve to literals of distinct nodes, so $G$ survives resolution.

\item \emph{Show that the common second gate is not pruned.}
Lemma~\ref{lem:pass}(c) gives $S_{z_1}\subseteq\{z_1,W_{z_1}\}$.
Use $Y=\{z_1\}$ when $W_{z_1}\notin S_{z_1}$, and $Y=\{z_1,W_{z_1}\}$ otherwise.
In the latter case, the only outgoing gate is $G$, which survives resolution and does not pass $z_1$.
If $W_{z_1}$ survives resolution, part (c) also gives $S_{W_{z_1}}=\{W_{z_1}\}$.
Part (b) places the new successors of $z_1$ within $\{W_{z_1},G\}$, and those of $W_{z_1}$ within $\{G\}$.

A surviving $W_{z_1}$ cannot be the new output.
That would place the original output in $S_{W_{z_1}}=\{W_{z_1}\}$.
But $W_{z_1}$ has successor $G$, whereas the original output has none.
Essentiality gives a path from $z_1$ to the new output.
The successor description forces it through $G$, so $G$ is not pruned.
The possible double charge from Step 7 is therefore excluded.

\item \emph{Exclude charges on gates that become constant.}
Every charged gate now has a role for a free input $z$.
Making $W_z$ or $W_z^2$ constant would cut every path from $z$ to the output.
This contradicts the essentiality of $z$ in $f|_\rho$.
\end{proofsteps}
\end{proof}

Elimination can create input triangles; the next lemma shows that normalization preserves the potential decrease.
\begin{lemma}[Normalization preserves the potential bound]\label{lem:trinorm2}
Suppose every gate of $C'$ is essential binary, has distinct predecessors, and reaches the output.
If its function has at least two inputs and satisfies (E1), then
$\Psi_2(N(C'))\le\Psi_2(C')$.
\end{lemma}
\begin{proof}
For $P=p(x,u)$ and $Q=q(x,P)$, replace $Q$ by $\psi(x,u)=q(x,p(x,u))$ as in Lemma~\ref{lem:normalize}.
Keep $P$ temporarily and replace the edge $P\to Q$ by $u\to Q$.
Since the table $\psi$ is nonlinear or degenerate, $7X+4A$ cannot increase.

A D-variable $z$ cannot equal $u$.
Its unique successor would be $P$, with two primary operands, contrary to Lemma~\ref{lem:forbid}(a).
Nor can $Q=W_z$: its primary operand $x$ feeds both $P$ and $Q$.
The changed edge and table therefore destroy no first credit.

Let $z$ have a second credit.
The added edge could alter $\succS(W_z)$ only if $u=W_z$.
Then $P=W_z^2$ would have primary operand $x$, contrary to Lemma~\ref{lem:roles}(a).
The removed edge could alter that successor set only if $P=W_z$.
This is impossible: $x$ has two successors and, as shown above, $u$ cannot be a D-variable.
Finally, $Q$ cannot have a second role because it has primary operand $x$.
Every second credit persists.

Apply Definition~\ref{def:resolve} with $I=\varnothing$, leaving the function unchanged.
To see that (E1), with at least two inputs, ensures dependence on every input, choose an input and fix a different one.
By (E1), changing only the chosen input changes the resulting function's output for some assignment.
Lemma~\ref{lem:charge2} assigns at most one lost unit to each eliminated nonlinear gate and none to degenerate gates.
Every gate contribution in equation~\eqref{eq:charge} is nonnegative, as is $\Gamma$.

The termination argument in Lemma~\ref{lem:normalize} applies to repeated triangle replacement and simplification.
Since no operation increases $\Psi_2$, the final normalized circuit has potential at most $\Psi_2(C')$.
\end{proof}

\section{Fourteen units per input}\label{sec:elimination}
We prove that an input can be fixed to decrease the potential by at least fourteen, refining Zwick's analysis of how many gates use each input \cite[Lemma~3.2]{Zwi91} for the gate types of Definition~\ref{def:type}.
In the tight case, eliminated gates contribute twelve units, while two surviving nonlinear gates form a single-successor chain that gives an input two new credits (Figure~\ref{fig:tight}).
This accounting applies to the entire circuit and permits arbitrary sharing.

\begin{lemma}[One step of elimination]\label{lem:mainB}
Let $C$ be normalized in the sense of Definition~\ref{def:norm}.
Suppose that it computes a function $f$ of at least two inputs satisfying (E2) and (S4).
Some primary input $x$ with $d(x)\ge2$ and some $c\in\bits$ satisfy
\begin{equation*}\tag{1b}\label{eq:elimination}
\Psi_2(C)-\Psi_2\bigl(N(C^{x=c})\bigr)\ge14.
\end{equation*}
\end{lemma}
The selected input has at least two successors, so it has no first credit before restriction.
Lemma~\ref{lem:charge2} therefore assigns at most one lost credit to each eliminated gate.

\begin{proof}
Inspect the first gate and its immediate successors.
When the eliminated gates do not already supply fourteen units, we use (S4) to separate the successor chains and examine the possible credit gains.
\begin{proofsteps}
\item \emph{Determine each gate's contribution.}\label{step:account}
Condition (S4) forces at least three inputs: on two inputs, the four pair cofactors would be Boolean constants and could not all differ.
By (E2), every one-input restriction leaves at least two essential inputs and satisfies (E1).
Lemma~\ref{lem:trinorm2} therefore lets us count the decrease before final normalization.

We will fix only an input of fan-out at least two, which cannot be a D-variable.
Lemma~\ref{lem:charge2} therefore assigns seven units to each eliminated parity gate and at least three to each eliminated nonlinear gate.
A nonlinear gate contributes four when it has no role or becomes constant.
Each newly gained credit adds one unit.

\item \emph{Locate the first gate and an input used at least twice.}\label{step:first}
Let $T$ be the first gate in topological order, with distinct primary operands $x,y$.
If both fed only $T$, all dependence on $(x,y)$ would pass through the single bit $T$.
The pair would have at most two cofactors, contrary to (S4).
Name the operands so that $d(x)\ge2$.

Two structural facts prevent double counting below.
A successor of a gate fed by a primary input $v$ cannot also be fed directly by $v$: that would be an input triangle.
Also, a gate made constant cannot be the output, because the cofactor has at least two essential inputs.
It has a successor in the original normalized circuit, and each such successor is eliminated.

Lemma~\ref{lem:roles} makes every direct successor of $x$ and every successor of $T$ role-free.
An eliminated nonlinear gate in either set therefore contributes four units.

\item \emph{Finish the case where $T$ is parity.}\label{step:parity}
If all successors of $x$ are parity gates, either assignment to $x$ eliminates at least two, giving fourteen units.
Otherwise, choose a nonlinear successor $P$ and block it through $x$.
The distinct gates $T,P$ contribute $7+4$, and a successor of the now constant $P$ is eliminated and contributes at least three.
This successor differs from $P$ by acyclicity and from $T$ because $T$ is first, giving a total of at least fourteen.
It remains to consider nonlinear $T$.

\item \emph{Reduce the nonlinear case to two successors per input.}\label{step:degree-two}
The input $y$ also has fan-out at least two, since otherwise blocking $T$ through $x$ would disconnect $y$, contradicting (E1) and Lemma~\ref{lem:disconnect}.

Let $c_T$ be the constant output of $T$ when blocked.
Write $\varepsilon,\delta$ for the blocking values of $x,y$ at $T$, respectively.
For $v=x$ or $v=y$, call fixing $v$ to its blocking value for $T$ the \emph{$v$-step}.
If $d(v)\ge3$, this step eliminates $T$, two other successors of $v$, and a successor of $T$.
These four gates are distinct: no successor of $T$ can also be a direct successor of $v$.
They each contribute at least four, for a total of at least sixteen.

We may therefore assume $d(x)=d(y)=2$ and write $\succS(x)=\{T,U\}$ and $\succS(y)=\{T,V\}$, where $U,V$ are role-free by Lemma~\ref{lem:roles}(a).

\item \emph{Exclude parity and matching blocking values at $U$ or $V$.}\label{step:matching}
If $U$ is parity, the $x$-step eliminates $T,U$ and a successor of $T$, contributing at least $4+7+4$.
These gates are distinct because the absence of input triangles prevents the successor of $T$ from being $U$.
If $V$ is parity, the $y$-step gives the corresponding estimate.

Suppose instead that the blocking values of $x$ at $T,U$ agree.
Fix $x$ to that value, making both gates constant.
Neither gate feeds the other, since both have primary operand $x$ and input triangles are absent.
If their successor sets have at least two distinct gates in total, choose one successor of $T$ and another gate in their union.
They are distinct from $T,U$ and contribute at least four and three, respectively.
Together with $T,U$, the total is at least fifteen.
If there is only one common successor, both its operands are constant, so it contributes four and has an eliminated successor contributing at least three.
This gives the same total.
Matching blocking values at $T,V$ are handled by fixing $y$ instead.

We are left with nonlinear $U,V$ whose blocking values oppose those of $T$ at $x,y$, respectively.

\item \emph{Reduce to a unique nonlinear successor of $T$.}\label{step:R}
The $x$-step eliminates $T,U$, contributing eight units.
If $T$ has two distinct successors, both are eliminated and each contributes at least four.
Neither equals $U$, because input triangles are absent, so the total is at least sixteen.

Otherwise, let $R$ be the unique successor of $T$; it differs from $U$ and $V$ because those gates have primary operands $x,y$.
If $R$ is parity, it contributes seven, bringing the total to fifteen.
If $R$ is nonlinear and $c_T$ blocks it, it becomes constant and contributes four.
Any successor of $R$ distinct from $U$ adds at least three.
If its only successor is $U$, then $U=u(x,R)$ has both operands constant.
An eliminated successor of $U$ adds the required three units.
These additional gates are distinct from the gates already counted by topological order.

The only remaining case has a unique nonlinear successor $R$ not blocked by $c_T$.
The $x$-step eliminates the three distinct, role-free gates $T,U,R$, which contribute twelve units.
The $y$-step likewise eliminates $T,V,R$ and contributes twelve.

\item \emph{Use (S4) to prove $U\ne V$.}\label{step:distinct}
We use the two-successor obstruction of Lachish and Raz~\cite[Lemma~3.8]{LR01}; its local argument applies because the gates in question are nonlinear.
Let $c_U,c_V$ be the outputs of $U,V$ when blocked.
If $U=V$, its two predecessors are $x,y$, and all dependence on this pair passes through $(T,U)$.
At the blocking value of $x$ for $T$, the first component is $c_T$ and the second ranges over both bits.
At the other value of $x$, the second component is $c_U$ and the first ranges over both bits.
These two sets share $(c_T,c_U)$, so their union has only three values.
The pair $(x,y)$ would therefore have at most three distinct cofactors, contradicting (S4).
Thus $T,U,V$ are distinct.

\item \emph{Classify the successors of $U$ and $V$.}\label{step:goodbad}
Neither $U$ nor $V$ is the output: if the output were $V$, blocking it through $y$ would make the whole cofactor constant, contradicting (E1).
The argument for $U$ uses $x$, and normalization of $C$ then ensures that both gates have successors.

Call $V$ \emph{good} when it has at least two successors, when its unique successor is parity, or when its unique successor is nonlinear and blocked by $c_V$.
Otherwise call it \emph{bad}.
A bad $V$ has one nonlinear successor $V_2$, and $c_V$ does not block that successor.
Use the same definitions for $U$, with $U_2$ its unique successor when bad.
This classification distinguishes the cases settled by gate removal from those in which new input credits may be needed.

\item \emph{Finish when $V$ is good.}\label{step:Vgood}
Fix $y$ to its blocking value for $V$, which is nonblocking for $T$.
The gate $V$ becomes constant and $T$ is bypassed; both contribute four.
If $V$ has at least two successors, two of them contribute at least $3+3$.
If its only successor is parity, that gate contributes seven.
In the remaining good case, its only successor is nonlinear and blocked by $c_V$; it becomes constant and contributes four, while an eliminated successor of that gate contributes at least three.
Together with the eight units from $T,V$, each case gives at least fourteen.
The counted descendants differ from $T,V$ by topological order.

\item \emph{Finish when $U$ is good.}\label{step:Ugood}
Fix $x$ to its blocking value for $U$.
The gate $U$ becomes constant and $T$ is bypassed, giving eight units.
The three alternatives in Step~\ref{step:Vgood}, with $U$ in place of $V$, give at least six more.
We may now assume both $U,V$ are bad.

\item \emph{Gain two credits when $V_2\notin\{R,U\}$.}\label{step:gain}
In the $x$-step, the gates $T,U,R$ contribute twelve.
If another gate is eliminated, it contributes at least three and the estimate follows.

Otherwise every other gate remains in $C^{x=\varepsilon}$.
In particular, $V$ remains: it differs from $U$ by Step~\ref{step:distinct}, and from $T,R$ by their positions and the absence of input triangles.
The gate $V_2$ also remains, since it is not $T$, which is first, and it differs from $U,R$ by the present assumption.

Neither gate of $\succS(y)=\{T,V\}$ passes $y$: $T$ is constant and $V$ survives resolution.
Lemma~\ref{lem:pass}(c) gives $S_y=\{y\}$.
As $V_2$ survives resolution and is the only successor of $V$, it also gives $S_V=\{V\}$.
Part (b) now shows that the new successor sets are exactly $\{V\}$ for $y$ and $\{V_2\}$ for $V$.

Both gates are nonlinear, and the input $y$ had two successors before restriction, so it gains a first and a second credit.
The potential decrease is therefore at least $12+2=14$.

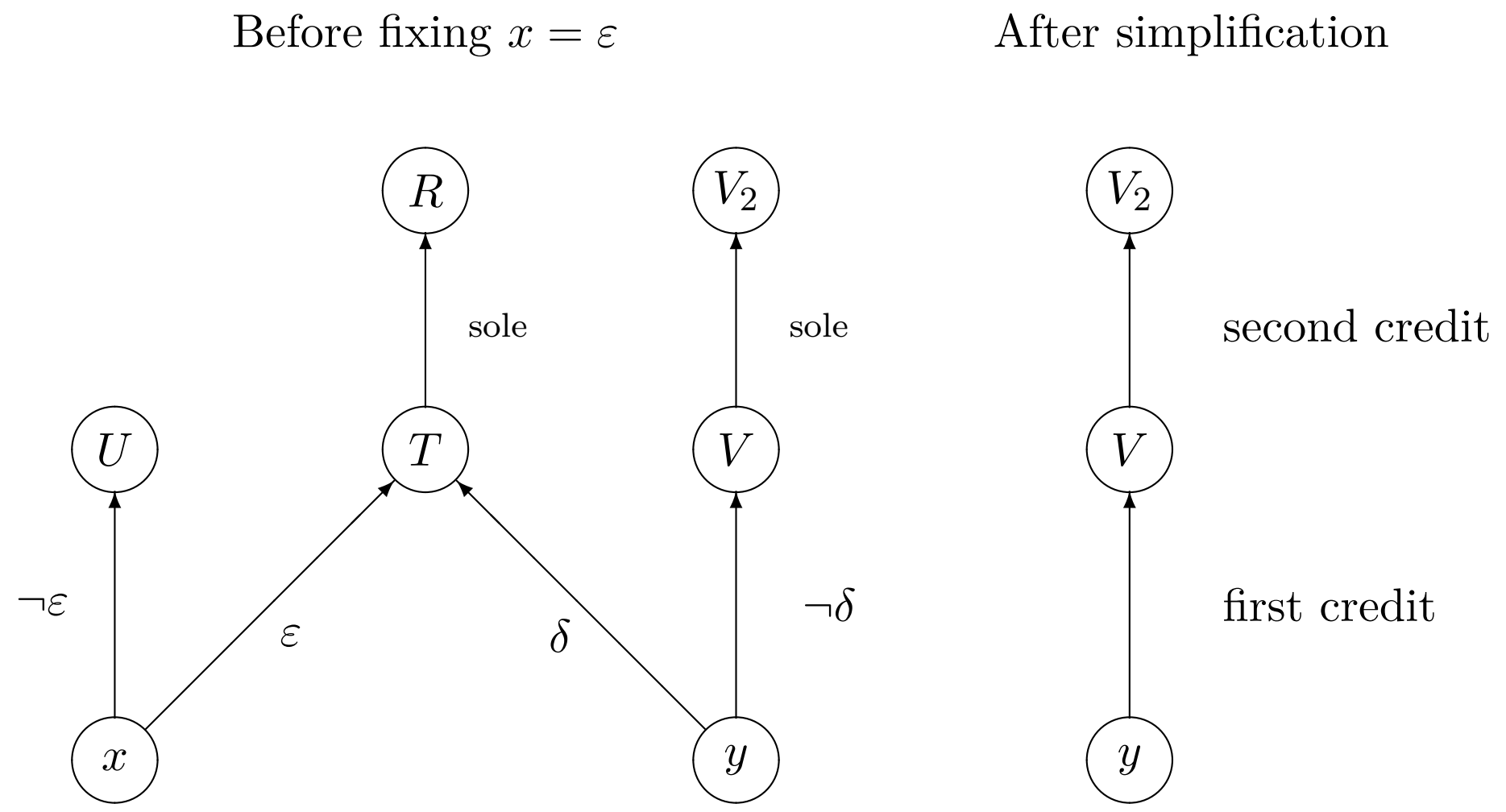
\begin{figure}[htbp]
\centering
\setlength{\unitlength}{0.85mm}
\begin{picture}(150,80)
\put(45,77){\makebox(0,0){Before fixing $x=\varepsilon$}}
\put(119,77){\makebox(0,0){After simplification}}
\put(15,7){\circle{8}}\put(15,7){\makebox(0,0){$x$}}
\put(75,7){\circle{8}}\put(75,7){\makebox(0,0){$y$}}
\put(15,37){\circle{8}}\put(15,37){\makebox(0,0){$U$}}
\put(45,37){\circle{8}}\put(45,37){\makebox(0,0){$T$}}
\put(75,37){\circle{8}}\put(75,37){\makebox(0,0){$V$}}
\put(45,62){\circle{8}}\put(45,62){\makebox(0,0){$R$}}
\put(75,62){\circle{8}}\put(75,62){\makebox(0,0){$V_2$}}
\put(15,11){\vector(0,1){22}}
\put(18,10){\vector(1,1){24}}
\put(72,10){\vector(-1,1){24}}
\put(75,11){\vector(0,1){22}}
\put(45,41){\vector(0,1){17}}
\put(75,41){\vector(0,1){17}}
\put(8,22){\makebox(0,0){$\neg\varepsilon$}}
\put(32,19){\makebox(0,0){$\varepsilon$}}
\put(58,19){\makebox(0,0){$\delta$}}
\put(84,22){\makebox(0,0){$\neg\delta$}}
\put(52,49){\makebox(0,0){\scriptsize sole}}
\put(83,49){\makebox(0,0){\scriptsize sole}}
\put(113,7){\circle{8}}\put(113,7){\makebox(0,0){$y$}}
\put(113,37){\circle{8}}\put(113,37){\makebox(0,0){$V$}}
\put(113,62){\circle{8}}\put(113,62){\makebox(0,0){$V_2$}}
\put(113,11){\vector(0,1){22}}
\put(113,41){\vector(0,1){17}}
\put(122,22){\makebox(0,0)[l]{first credit}}
\put(122,49){\makebox(0,0)[l]{second credit}}
\end{picture}
\caption{The tight case of Lemma~\ref{lem:mainB}, Step~\ref{step:gain}, when only $T,U,R$ are eliminated and $V_2\notin\{R,U\}$.
Edge labels give the blocking values at the receiving gate; other operands and edges are omitted.
The marked successors of $T$ and $V$ are unique.
After restriction, $y$ and $V$ each have the single successor shown, and both $V,V_2$ are nonlinear.
The removed gates give twelve units; the two new credits give the remaining two.}
\label{fig:tight}
\end{figure}

\item \emph{Prune the unused gate when $V_2=U$.}\label{step:prune}
Fix $x$ to its blocking value for $U$.
Then $U$ becomes constant and $T$ is bypassed, contributing eight.
If $V$ survives resolution, its only successor is the constant gate $U$.
Lemma~\ref{lem:pass}(c) gives $S_V=\{V\}$, and part (b) leaves it with no successor in the simplified circuit.

It cannot be the new output: the original output is not $V$, since $V$ has successor $U$, and no other node belongs to $S_V$.
Thus $V$ is pruned.
Whether removed by resolution or by pruning, it contributes four because it is role-free.
Finally, $U_2$ is eliminated and contributes at least three.
It differs from $V$, which precedes $U$, and from $T,U$ by topological order.
The total is at least fifteen.

\item \emph{Gain two credits in the reversed configuration.}\label{step:gain-reverse}
Suppose $V_2=R$ and $U_2\notin\{R,V\}$.
In the $y$-step, the role-free gates $T,V,R$ contribute twelve.
If any further gate is eliminated, the total is at least fifteen.
Otherwise $U,U_2$ remain.
Lemma~\ref{lem:pass} gives $S_x=\{x\}$ and $S_U=\{U\}$, exactly as in Step~\ref{step:gain}.
The new successor sets are $\{U\}$ for $x$ and $\{U_2\}$ for $U$.
The input $x$ gains both credits, giving fourteen units.

\item \emph{Exclude $V_2=R$ and $U_2=V$.}\label{step:opposite}
Let $\varepsilon$ be the blocking value of $x$ for $T$, and $\delta$ that of $y$.
Fix $x=\neg\varepsilon$, which blocks $U$.
Because $U$ is bad, its constant value $c_U$ does not block $V=U_2$.
The gate $V$ therefore becomes a literal of $y$.
It equals $c_V$ exactly when $y=\neg\delta$, the blocking value for $V$.
The gate $T$ also becomes a literal of $y$, equal to $c_T$ exactly when $y=\delta$.

The gate $R$ has operands $T,V$.
Neither $c_T$ nor $c_V$ blocks it.
Its blocking values are consequently $\neg c_T$ at $T$ and $\neg c_V$ at $V$.
When $y=\delta$, its $V$-operand is $\neg c_V$ and blocks it.
When $y=\neg\delta$, its $T$-operand is $\neg c_T$ and blocks it.
Both blocking assignments give the same constant output of the nonlinear gate $R$.

Every path from $y$ begins through $T$ or $V$, both of which feed only $R$.
Thus the cofactor is independent of $y$, contradicting (E1).

\item \emph{Handle the remaining case.}\label{step:remaining}
The preceding steps cover every case except $V_2=R$ and $U_2=R$.
In that case $R$ would have the three distinct operands $T,U,V$, which is impossible for a binary gate.
\end{proofsteps}
In each case the selected input has fan-out at least two and the decrease is at least fourteen.
Final normalization cannot increase the potential by Step~\ref{step:account}, proving equation~\eqref{eq:elimination}.
\end{proof}

Table~\ref{tab:local-cases} collects the final alternatives and their estimates.
\begin{table}[htbp]
\centering\small
\begin{tabular}{@{}p{0.35\linewidth}p{0.20\linewidth}p{0.35\linewidth}@{}}
\toprule
Condition & Restriction & Potential decrease or exclusion\\
\midrule
$V_2\notin\{R,U\}$ & $x=\varepsilon$ & At least $12+2$; Step~\ref{step:gain}\\[3pt]
$V_2=U$ & $x=\neg\varepsilon$ & At least $15$; Step~\ref{step:prune}\\[3pt]
$V_2=R$, $U_2\notin\{R,V\}$ & $y=\delta$ & At least $12+2$; Step~\ref{step:gain-reverse}\\[3pt]
$V_2=R$, $U_2=V$ & $x=\neg\varepsilon$ & Contradicts (E1); Step~\ref{step:opposite}\\[3pt]
$V_2=R$, $U_2=R$ & None & Three operands at $R$; Step~\ref{step:remaining}\\
\bottomrule
\end{tabular}
\caption{The final alternatives when both $U,V$ are bad in Lemma~\ref{lem:mainB}.
The blocking values $\varepsilon,\delta$ at $T$ are defined in Step~\ref{step:degree-two}.
In either $12+2$ row, two credits supply the last two units if only three gates disappear; any additional eliminated gate supplies at least three.}
\label{tab:local-cases}
\end{table}

\section{The gate count for 3-SAT}\label{sec:main}
To bound the circuit size of the combined function, we sum the potential decreases as the whole circuit is restricted, counting one-bit steps for each block until it reaches $k+1$ free inputs.
We then add the nonlinear-gate count supplied by full degree.
\begin{theorem}[Composition of robust blocks]\label{thm:compB}
Let $G$ be a read-once composition whose $B$ disjoint blocks are restrictions of $k$-robust functions.
Every block retains at least $k+2$ free inputs.
Write $L$ for their total number, and regard circuits for $G$ as having exactly these $L$ primary inputs.
Every normalized circuit computing $G$ satisfies
\begin{equation}\label{eq:treepotential}
\Psi_2(C)\ge14\bigl(L-(k+1)B\bigr).
\end{equation}
If all the blocks have full degree, their composition satisfies
\begin{equation}\label{eq:treesize}
\C(G)\ge\left\lceil\frac{17L-14(k+1)B-3}{7}\right\rceil.
\end{equation}
\end{theorem}
Full degree in the second assertion is required of the blocks after their stated restrictions, on the inputs that remain free.
The term $L-(k+1)B$ in~\eqref{eq:treepotential} counts the one-bit eliminations to which Lemma~\ref{lem:mainB} applies, while the extra $3L-3$ in~\eqref{eq:treesize} comes from the nonlinear gates forced by full degree.
\begin{proof}
We alternate a one-bit elimination with removal of any block that has reached the threshold.
The latter operation does not increase the potential, so we count only the one-bit steps.
\begin{proofsteps}
\item \emph{Eliminate one bit while all blocks are long enough.}
If each block has at least $k+2$ bits, Lemma~\ref{lem:separate} gives (E2) and (S4) for the composition.
Apply Lemma~\ref{lem:mainB} and normalize after its restriction.
The potential decreases by at least fourteen, and exactly one block loses one free input.
\item \emph{Remove a block at the threshold.}
When a block first reaches $k+1$ bits, it is still nonconstant by robustness.
If another block remains, assign all bits of the short block to make its value one at an AND parent or zero at an OR parent.
Its sibling then replaces the parent in the defining tree.
Before this restriction, every leaf has at least $k+1$ bits, so Lemma~\ref{lem:separate} gives (E1).
Afterward, the remaining composition depends on every free input and still satisfies (E1).
The charging identity of Lemma~\ref{lem:charge2} has nonnegative contributions: even a twice-charged nonlinear gate contributes $4-2=2$.
Resolution therefore cannot increase $\Psi_2$.
Lemma~\ref{lem:trinorm2} shows the same for subsequent normalization.
If only one block remains, stop when it reaches $k+1$ bits.
\item \emph{Sum the decreases.}
A block of initial length $\ell$ undergoes exactly $\ell-(k+1)$ one-bit steps before removal or the final stop.
Hence the total number of such steps is $L-(k+1)B$.
In the terminal circuit, $D^*\le D\le2A$, where the second inequality holds because each first credit uses one of the two operands of a nonlinear gate.
Thus $D+D^*\le4A$, and equation~\eqref{eq:psi} gives $\Psi_2\ge0$.
Each counted step decreases the potential by at least fourteen, while every other operation has nonnegative decrease.
This proves~\eqref{eq:treepotential}.
\item \emph{Convert the potential to circuit size.}
Normalize a minimum circuit for $G$ by Lemma~\ref{lem:normalize}.
Every gate is parity or nonlinear, so equation~\eqref{eq:psi} rearranges to
\[
7|C|=\Psi_2(C)+3A(C)+D(C)+D^*(C).
\]
Full degree and Proposition~\ref{cor:degL} give $A(C)\ge L-1$; substituting this bound and using the nonnegativity of the two credit counts yields
\[
7|C|\ge14\bigl(L-(k+1)B\bigr)+3(L-1)
=17L-14(k+1)B-3.
\]
Since $|C|$ is an integer, dividing by seven and taking the ceiling proves~\eqref{eq:treesize}.
\end{proofsteps}
\end{proof}

We now obtain Theorem~\ref{thm:overview}, together with its finite form, by embedding these blocks in variable identifiers and applying the transfer lemma.
\begin{theorem}[Gate complexity of binary-encoded 3-SAT]\label{thm:main}
Let $F_w$ be the decision function of Definition~\ref{def:F} for exactly $m=2^w-1$ clauses over $m$ available variables.
Its encoding has $n=3m(w+1)$ bits.
For unrestricted binary circuit complexity, along these input lengths as $w\to\infty$,
\begin{equation}\label{eq:main}
\C(F_w)\ge\frac{17}{7}n-O\!\left(\frac{n\log\log n}{\log n}\right).
\end{equation}
More precisely, let integers $r,k$ satisfy
$4\le r\le w-1$, $3\le k\le r-2$, and
$2^{k-2}>\log_2(6r(r-1))+r\log_2 3$.
Then
\begin{equation}\label{eq:finite}
\C(F_w)\ge n+\left\lceil\frac{(30r-42(k+1))(m-2^r)-3}{7}\right\rceil.
\end{equation}
A restriction using the weighted modular test of Proposition~\ref{prop:explicit} gives
$\C(F_w)\ge(17/7)n-O(n/\sqrt{\log n})$.
\end{theorem}
\begin{proof}
We obtain a finite bound from the number of positive clauses left by the embedding, bound that number independently of the chosen test, and choose parameters giving a sublinear error.
\begin{proofsteps}
\item \emph{Build the restricted SAT function.}
Choose a full-degree $k$-robust test $g$ on $r$ bits using Proposition~\ref{prop:random}.
Lemma~\ref{lem:embedg} leaves $M=m-|g^{-1}(0)|$ positive clauses, giving $B=3M$ disjoint blocks and $L=3Mr$ free bits.
Associate the three terms in each clause using two binary ORs, and combine the clauses using a binary AND tree.
The condition $r\ge k+2$ lets Lemma~\ref{lem:separate} verify that every free bit is essential.
\item \emph{Transfer the composition bound.}
Lemma~\ref{lem:transfer} adds the $n-L$ removed gates to~\eqref{eq:treesize}.
Substituting $L=3Mr$ and $B=3M$ gives
\begin{equation}\label{eq:transfer}
\begin{aligned}
\C(F_w)
&\ge n-L+\left\lceil\frac{17L-14(k+1)B-3}{7}\right\rceil\\
&=n+\left\lceil\frac{(30r-42(k+1))M-3}{7}\right\rceil.
\end{aligned}
\end{equation}
\item \emph{Derive the uniform finite inequality.}
Since $|g^{-1}(0)|\le2^r$, we have $M\ge m-2^r$.
If $30r-42(k+1)\ge0$, substitute this lower bound for $M$ in~\eqref{eq:transfer} to obtain~\eqref{eq:finite}.
If that coefficient is negative, it is at most $-6$, because it is a multiple of six.
Also $r\le w-1$ implies $m-2^r\ge1$.
The fraction inside the ceiling in~\eqref{eq:finite} is then at most $-9/7$, whose ceiling is $-1$.
In this case the elementary bound $\C(F_w)\ge n-1$ from Lemmas~\ref{lem:floor} and~\ref{lem:essential} suffices.
\item \emph{Retain almost all description bits.}
Take $r=w-\lceil\log_2 w\rceil-1$.
Then
\[
2^r\le\frac{2^w}{2w}=\frac{m+1}{2w}\le\frac{m}{w},
\qquad M=m-O(m/w).
\]
Using $n=3m(w+1)$ and $L=3Mr$, the number of fixed bits is
\[
n-L=3m(w+1-r)+3(m-M)r=O(m\log w)
=O(n\log w/w).
\]
The first term counts the high identifier and sign bits fixed in every literal field; the second counts the additional low identifier bits fixed in the helper clauses.
\item \emph{Choose the logarithmic threshold and estimate the loss.}
The right side of the counting hypothesis in Proposition~\ref{prop:random} is of order $r$.
The least admissible integer $k$ is therefore $\log_2 r+O(1)$.
For sufficiently large $w$, we have $r\ge k+2$ and all finite hypotheses hold.
Since a ceiling is at least its argument, the first line of~\eqref{eq:transfer} gives
\[
\C(F_w)\ge\frac{17}{7}n
-\frac{10}{7}(n-L)-2(k+1)B-\frac37.
\]
Now $B=3M\le3m=n/(w+1)$.
Step~4 and $k=O(\log w)$ make the total loss $O(n\log w/w)$.
Finally $\log_2 n=w+O(\log w)$, so this is $O(n\log\log n/\log n)$.
This proves~\eqref{eq:main}.
\item \emph{Use the weighted modular test.}
Keep the same choice of $r$, but use Proposition~\ref{prop:explicit}, whose threshold is $k=O(\sqrt r)$.
For large $w$, the requirement $r\ge k+2$ again holds.
The derivation of~\eqref{eq:transfer} used only robustness and full degree, so it applies to this test too.
The loss is now
\[
O(n\log w/w+nk/w)=O(n/\sqrt{\log n}).
\]
The residue selected in Proposition~\ref{prop:explicit} specifies the forbidden identifiers in Lemma~\ref{lem:embedg}.
\end{proofsteps}
\end{proof}

\subsection{Formalization and reproducibility}\label{sec:formalization}
The Lean proofs are available on Zenodo~\cite{Companion}.
The accompanying supplement provides instructions for reproducing the verification and a correspondence between the manuscript's results and the formalized theorems.

The Lean evaluator for formulas implements Definition~\ref{def:F}, including rejection of identifier zero.
Circuits are represented both by ordered lists of gates, each referring only to earlier signals, and by acyclic graphs with arbitrary vertex labels, shared predecessors and a designated output.
Unary and constant gates use tables that ignore their unused arguments, which can be filled with any primary input without adding gates or cycles.
The two representations compute the same function, and the theorems for the graph model give the finite bound~\eqref{eq:finite} and both asymptotic estimates under the hypotheses of Theorem~\ref{thm:main}.

The gate-elimination proof records which inputs remain free and requires the circuit to use no inputs outside that set when applying its potential bounds, as in Theorem~\ref{thm:compB}.
The implementation counts occurrences of signals as gate arguments or as the output; separate lemmas show that its credit counts agree with Definitions~\ref{def:D} and~\ref{def:star} for normalized circuits.

The weighted modular construction is implemented using lists of Boolean values and unary counters, with one symbol per counted unit.
For this representation, the formalization proves a polynomial bound on the number of elementary operations needed to choose the prime and find the residue.
The $O(rp)$ bound in Proposition~\ref{prop:explicit} instead counts the residue computation once the prime is supplied.

The development uses Lean 4.35.0 and a fixed revision of the mathlib library, and the audit checks that the proofs use only the standard Lean axioms \texttt{propext}, \texttt{Classical.choice} and \texttt{Quot.sound}.

\section{Conclusion}\label{sec:conclusion}
Theorem~\ref{thm:main} gives a lower bound of
$(17/7)n-O(n\log\log n/\log n)$ gates for the binary SAT family of Definition~\ref{def:F}, counting every bit of the formula description and permitting arbitrary binary gates, depth and sharing.

The embedding places Boolean tests in variable identifiers and leaves $n-o(n)$ description bits free; the weighted gate count applies even when circuits share gates across tests.
When gate removal alone gives only twelve units of potential, new credits supply the two units needed to reach fourteen.
The nonlinear gates required by full degree then yield the coefficient $17/7$.

Further work could seek explicit tests that remain suitable for gate elimination after more inputs are fixed, reducing the $O(n/\sqrt{\log n})$ loss.
Varying the ratio of clauses to available variables may offer embeddings that preserve the identifier tests while retaining nearly all description bits.
Constraints from repeated variable occurrences could also strengthen the gate count.
The formalized finite inequality provides a baseline for assessing these improvements.

\paragraph{Author contributions.}
The authors contributed equally and share first authorship.
\paragraph{AI disclosure.}
AI tools (Claude Code, Codex) assisted with manuscript structure, literature search, Lean verification, reference checks, and LaTeX and proof formatting.
The proof ideas were developed by the authors, who take full responsibility for the manuscript's content, mathematical arguments, and results.
\bibliographystyle{unsrt}
\bibliography{references}
\end{document}